# Method of "active correlations" for DSSSD detector application

*Yu.S. Tsyganov*

FLNR, JINR 141980 Dubna, Russia

tyura@sungns.jinr.ru

**Abstract**

*Real-time PC based algorithm is developed for **DSSSD** ( **D**ouble **S**ide **S**ilicon **S**trip **D**etector)  detector. Brief description of the detection system is also presented. Complete fusion nuclear reaction $^{nat}Yt+^{48}Ca \rightarrow {}^{217}Th$ is used to test this algorithm at $^{48}Ca$ beam. Example of successful application of a former algorithm for resistive strip **PIPS** (resistive strip **P**assivated **I**mplanted **P**lanar **S**ilicon) detector is presented too.*

1. Introduction

The existence of superheavy elements (SHE) was predicted in the late 1960s as one of the first outcomes of the macroscopic-microscopic theory of atomic nucleus. Modern theoretical approaches confirm this concept. To date, nuclei associated with the "island of stability" can be accessed preferentially in $^{48}$Ca-induced complete fusion nuclear reactions with actinide targets. Successful use of these reactions was pioneered employing the Dubna Gas-Filled Recoil Separator (DGFRS) at the Flerov Laboratory of Nuclear Reactions (FLNR) in Dubna, Russia [1]. In the last two decades intense research in SHE synthesis has taken place and lead to significant progress in methods of detecting rare alpha decays. Method of "active correlations" [2] used to provide a deep suppression of background products is one of them. Significant progress in the detection technique was achieved with application of DSSSD detectors. In particular, DSSSD detector was applied in the experiment with the reaction $^{249}Bk+^{48}Ca \rightarrow 117+3,4n$ [3] aimed at the synthesis of Z=117 element. Note that applying the method of "active correlations" with DSSSD detector is even more effective compared with the case of resistive strip PIPS detector. On the other hand, some specific effects take place and possible sharing registered signal between two neighbor strips is one of them. The aim of this paper is to present development of method of "active correlations" for application with DSSSD detector.

2. Detector types for SHE research

There are two main types of silicon radiation detectors for SHE research:

1. Detectors working on the principle of charge dividing (resistive strip PIPS detectors) ;
2. Detectors working on the principle of creating of effective X-Y position cell as a crossing of front and back strip lines (double side silicon strip detector (DSSSD)).

In case of PIPS detector the main idea of the "active correlations" method is aimed at searching in real-time mode of time-energy-position recoil-alpha links (ER-α), using the discrete representation of the resistive layer of PIPS detector. Elapsed ER's time is used as a recoil matrix element stored in

PC's RAM. The block diagram of the process is shown in the Fig.1. Note that in contrast to the real-time algorithm described in [4] two recoil matrixes are used to store the value of elapsed ER time signal (matrixes "top" and "bottom", respectively).

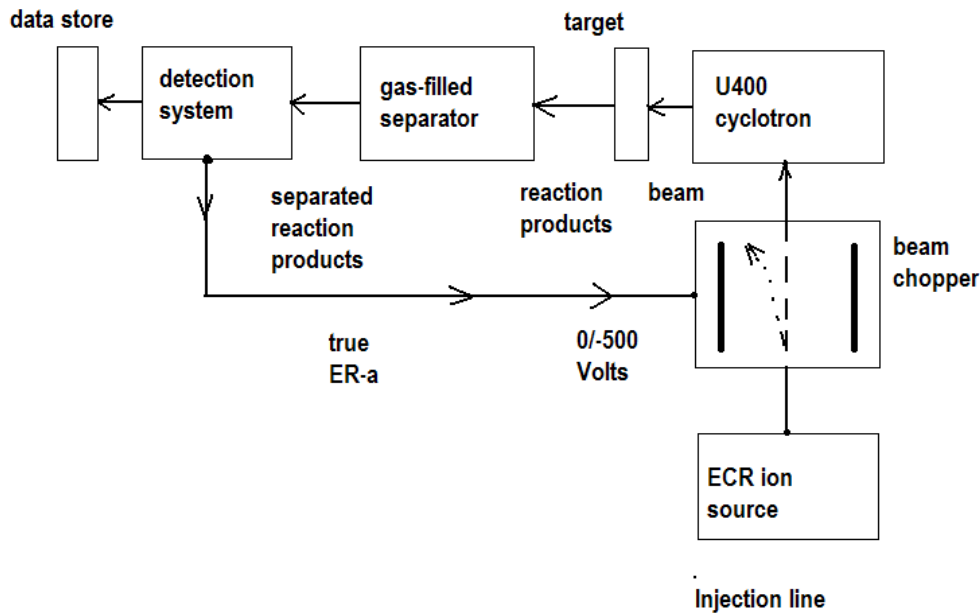

Fig.1 The block diagram of the process. Dead time is about: 35 µs (electronics) + 60 µs (an orbit life time)

As an example, in Fig.2 decay of Z=117 element is presented.

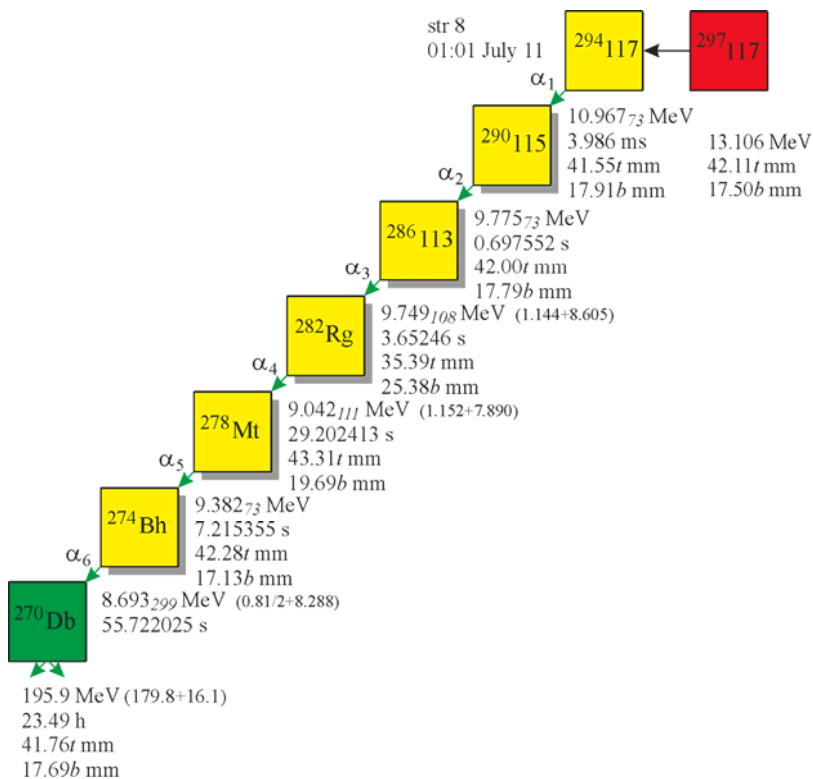

Fig.2 Z=117 decay chains detected in strip #8 of 32 strip PIPS (CANBERRA NV) detector. Shadows indicate beam-off phase.

## 3. Detection module and CAMAC electronics

Attractive features of DSSSD detectors such as good energy and position (X, Y) resolution are well known. The necessity to use more electronics channels is surpassed with modern CAMAC electronic modules manufactured by *Techinvest* group [5]. Together with standard charge sensitive preamplifiers it takes only two crates to process signals from 48x128 strip DSSSD. Additionally, six signals from backward detectors (with no position sensitivity) are measured. Two low pressure pentane filled multiwire proportional chambers (MWPC) START and STOP are used to provide both time-of-flight (TOF) and "ΔE"(start/stop) information about charge particle coming from cyclotron. In Table.1 main modules are presented with brief specification of their functions.

**Table 1. Main modules of the detection system**

| Module | Function | Self triggering | Reset | comments |
|---|---|---|---|---|
| PA-27n ADC (16 inputs) | Signals : 3x16 (48 front strips)+8x16 back strips + 1x6 backward detector | No, triggering-fast signal from 16 in shaping amplifier | F0/2 and F10 CAMAC functions | 12 (1 M), 11 bit 2 μs conv. time |
| PA-25 ADC | TOF, ΔE1, ΔE2 from MWPC START/STOP | yes | F0/2 , F10 and 20-50 μs after Q=1 if no F0/2 F28 – to write state bits | 3 (1M), 10 bit 2 μs conv. time |
| D-16 universal module (state register) | 16 state bits (1/0), each one shows TTL level; + test pulser 1μs 600 Hz | No, F28 CAMAC function is required | F0/2, 10 Change of TTL levels of additional outputs: NA(7,8) F16 0/1 | 2M |
| Timer TechInvest | | Elapsed time – two words, 16 bit each. First word – 1 μs. Overflow - +1 count to another word. Additionally word for synchronization time (1μs)- for synchronization between ADC's | | 1M |
| CC 212 M | Crate controller | - | - | 2M ( PCI card) |
| Misitec DDD (standard) | Charge sensitive preamplifier | | | |
| Detector | 48 x 128 strips | - | - | Micron Scd. (UK) |
| Twin fast preamplifier for START/STOP signals from MWPC's | | | | Custom made |

## 4. Specific of DSSSD detector application

As to the specificity of applying DSSSD detector and development of a corresponding real-time algorithm, one should keep in mind the following:

- better position resolution in comparison with PIPS detector;

- detector's structure (XY strips 48x128) corresponds to matrix of the given dimensions which, in first approximation, can be used as the matrix of recoil nuclei; its elements are filled in by value of the current time taken from CAMAC hardware upon receiving the corresponding events;

- due to presence of P+ isolating layer between two neighbor strips on the ohmic side of the detector (48 front strips) the edge effects are negligible;

- on the contrary, for the 128 back strips (p-n junction) the effect of charge sharing between neighboring strips can be up to some 17% in the geometry close to $2\pi$ [6]. Certainly, this effect should be taken into account when developing and applying the algorithm of search of potential ER-α correlation. In addition to Table 1 indicating main electronic modules, below is presented the 14-word (16 bit each) event specification as C++ code fragment.

In Fig.3a operation bits of the position code of the state register are shown. These codes correspond to $2^n$ (n=0...3) value indicate single back strip signal, whereas $2^n+2^m$ codes (m=0...3) indicate detection of signals from two neighbor strips (Fig.3b).

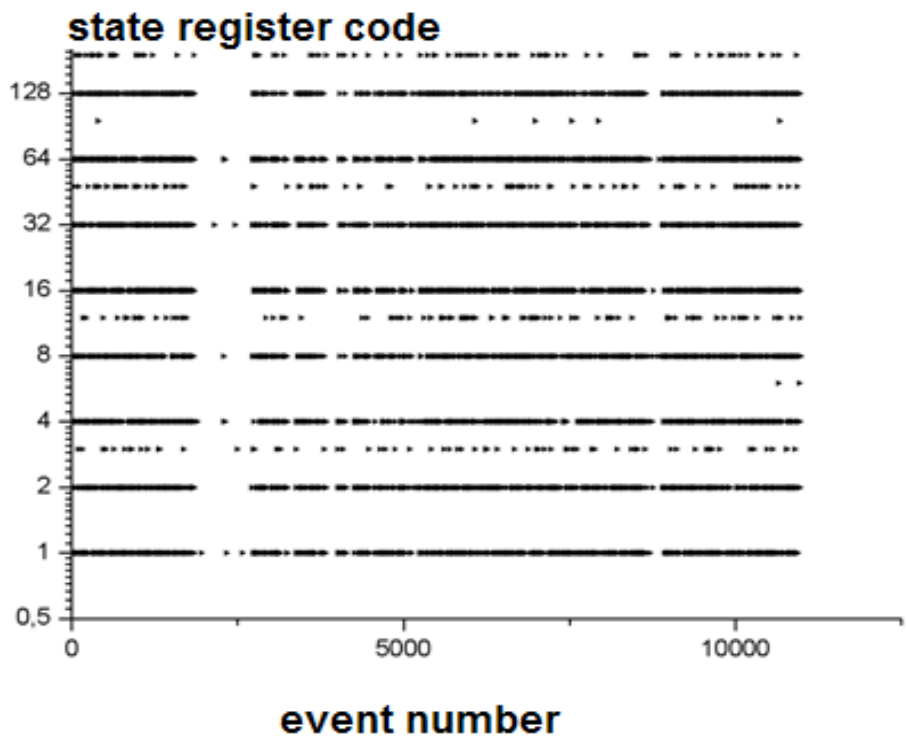

Fig. 3a State register codes. ( bit's 1 to 8 correspond to eight back strip 16 in ADC's).

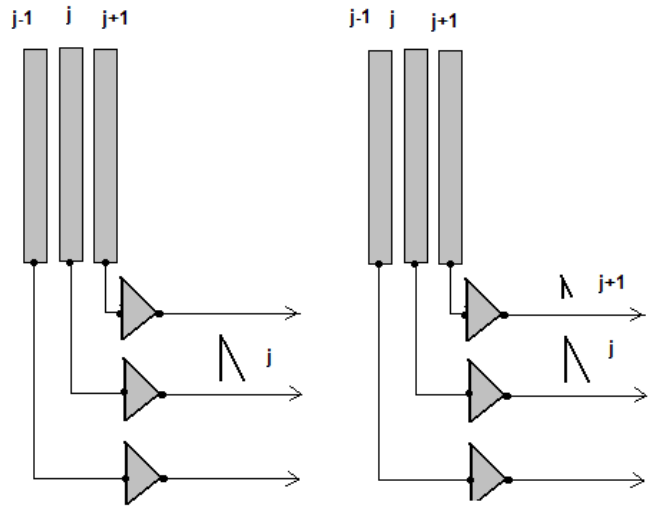

**Fig.3b** Two situations for back strips j, j+/-1 with output preamplifier signals shown schematically (single $2^n$ and shared with two neighbor strips as $2^n+2^m$)

```
// --- C++ code fragment – one event from DSSSD

    ID_EVN+(beam_marker<<4)+(nm_adcb_1<<8)+(nm_adcb_2 << 12 );

    buffer[WORDS_IN_EVENT*cnt_evn+1 ] = wa;

    buffer[WORDS_IN_EVENT*cnt_evn+2 ] = wf;

    buffer[WORDS_IN_EVENT*cnt_evn+3 ] = t_high;

    buffer[WORDS_IN_EVENT*cnt_evn+4 ] = t_low;

    buffer[WORDS_IN_EVENT*cnt_evn+5 ] = t_mcs;

    buffer[WORDS_IN_EVENT*cnt_evn+6 ] = tof;

    buffer[WORDS_IN_EVENT*cnt_evn+7 ] = wab;

    buffer[WORDS_IN_EVENT*cnt_evn+8 ] = wfb;

    buffer[WORDS_IN_EVENT*cnt_evn+9 ] = wabb;

    buffer[WORDS_IN_EVENT*cnt_evn+10] = wfbb;

    buffer[WORDS_IN_EVENT*cnt_evn+11] = t_sinhro;

    buffer[WORDS_IN_EVENT*cnt_evn+12] = DE1;

    buffer[WORDS_IN_EVENT*cnt_evn+13] = DE2;
```

The values which compose one event are:

- ID_EVN's bits 1 to4: 0..3 – code of an actual ADC (front strip 1..48 -0,1,2 or backward detector ADC if ID_EVN=3), ID_EVN's bits 5 to 8 – beam marker (0/1), bits 9 to 12 and 13 to 16 – numbers of back strip ADC's;
- wa (13 bit) – code of amplitude ($\alpha$ – scale), wf bits 1 to 12 - code of amplitude fission fragment scale, bits 13 to 16 code of ADC channel (1..16) ;

- t_high – elapsed time counter1;
- t_low – elapsed tome counter2, elapsed time value is t=65536*t_high + t_low in microseconds;
- t_mcs – relative time in microseconds, reset each 32 ms by the signal from the pair photodiode/light emitting diode of the rotating target wheel;
- tof (12 bit) – time of flight amplitude;
- wab – the same as wa, but for back strip ADC;
- wfb – the same as wf;
- wabb – the same as wab, but it corresponds to the signal from neighbor strip in the case of signal sharing;
- wfbb – the same as wfb;
- t_sincho – time value (in µs) for event synchronization when two ADC's (front and backward ) are actual;
- $DE_{1,2}$ – codes (12 bit) of amplitude for signals from START and STOP proportional chambers, respectively;
- Words_in_event = 14, and cnt_evn – event counter (1...512), buffer is an array for data writing process.

For calibration complete is used complete-fusion reaction $^{nat}$Yt+$^{48}$Ca→Th+xn. Usually, least squares using eight peaks method is applied to extract linear calibration constants $a_i$ and $b_i$ and $a_j$ and $b_j$ for 48 and 128 strips respectively (i=1...48; j=1...128). The measured energy in this case is calculated as $E_i=a_i+b_i*N$, where N- spectrum channel number. The flowchart of C++ code to search for ER-α correlation is shown in the Fig.4 a,b. Note, that when filling ER matrix element both neighbor back strips are also filled with the same elapsed time values.

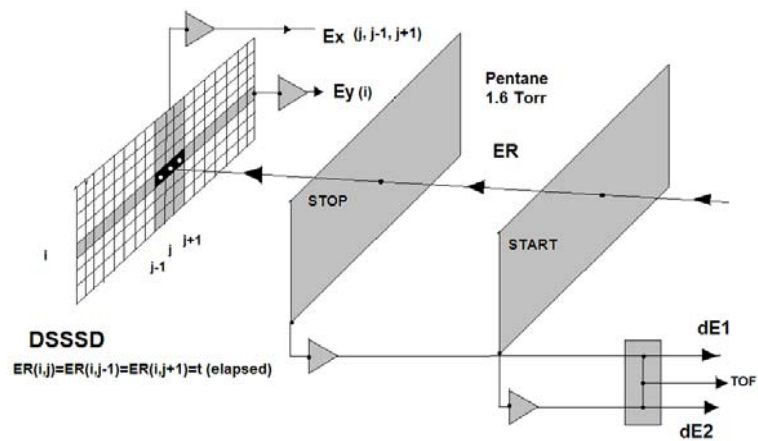

Fig. 4 a) The schematic of the ER-matrix formation process. Proportional chambers START, STOP and DSSSD detector are shown.

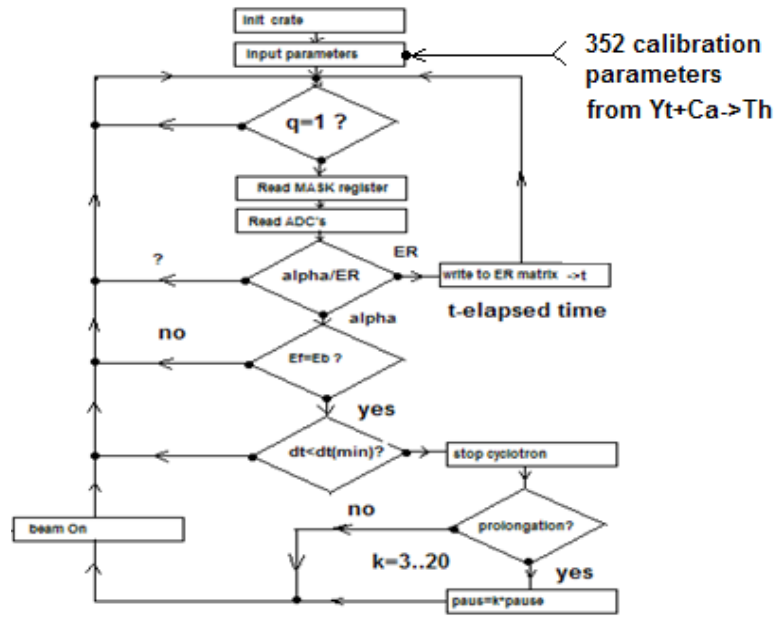

Fig. 4b) The flowchart of the process of ER-alpha chains search . $E_f$-front strip energy, $E_b$-back strip energy; MASK register (see. Tab.1) allows establishing which ADC's are operated.

## 5. Test in $^{nat}$Yt+ $^{48}$Ca →$^{217}$Th+3n complete fusion nuclear reaction

The algorithm of search for ER-alpha correlations was tested in two stages. In the first test $^{238}$Pu was used as alpha-particle source; the data acquisition program generated time-of-flight signal for a certain part of events, using the simple formula (fragment coded in C++) :   ntt = random (100); if (ntt >30 )  tof =(TOF_MIN+TOF_MAX)/2-random(600)+random(550); else tof=0;  if (ER==true). The energy threshold for alpha-particles as well as for recoil nuclei was set below 5 MeV. All the found ER-alpha correlations were written into file with exact value of file time. Further the data was processed off line. Debugging of program was considered complete when reaching unambiguous correspondence of the observed "correlations". The principal test was run with the reaction $^{nat}$Yt + $^{48}$Ca→$^{217}$Th+3n after calibrating both front and back strips in this reaction. For the correlation chains found in the time window of 0-2$T_{1/2}$ ($^{217}$Th) and energy window of 9261+/-30 KeV the energies of $^{217}$Th recoils were measured. The corresponding energy spectrum is shown in Fig.5 a,b.

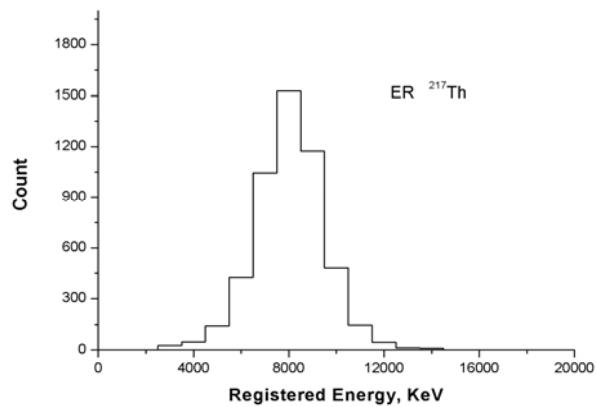

Fig.5a  Spectrum of registered energy of $^{217}$Th recoils.  The mean value is equal to 7563; std.= 1477; *The extracted time interval $\Delta t = 0,475\ ms$*

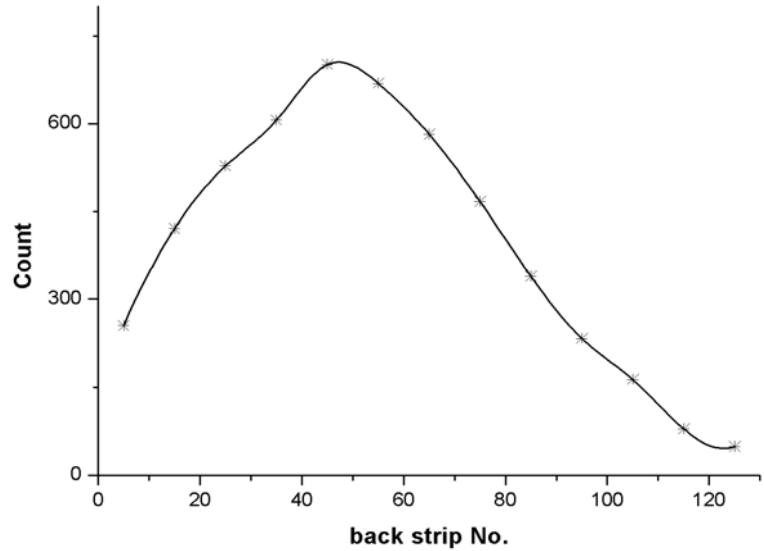

Fig.5 b Distribution of extracted events versus to back strip number (line corresponds to a cubical spline )

Note, that events are grouped in the detector center, whereas the shape of the ER measured energy spectra corresponds well with the calculations [6].

## 6. Conclusions

- Method of "active" correlations was successfully applied in the $^{249}$Bk+$^{48}$Ca→$^{294,293}$117+3,4n experiment when using resistive layer position sensitive PIPS detector;

- Real-time method for DSSSD detector to suppress beam associated background products in heavy-ion-induced complete fusion nuclear reactions was implemented and tested in $^{nat}$Yt+ $^{48}$Ca→$^{217}$Th+3n reaction;

- Neighbor strips edge effects are taken into account in the algorithm development;

- This method will be applied in the long term experiments aimed to the synthesis of new superheavy isotopes.

## 7. Acknowledgments

This paper is supported in part by the RFBR Grant №13-02-12052.
Author is grateful to Drs. A.N.Polyakov, A.Voinov , A.Sukhov and V.Zhuchko for their continuous support in many years of collaboration.